\documentclass[conference]{IEEEtran}
\usepackage{listings}

\usepackage{cite}
\usepackage{amsmath,amssymb,amsfonts}
\usepackage{algorithmic}
\usepackage{graphicx}
\usepackage{textcomp}
\usepackage{url} 
\usepackage{xcolor}
\usepackage{hyperref}

\def\BibTeX{{\rm B\kern-.05em{\sc i\kern-.025em b}\kern-.08em
    T\kern-.1667em\lower.7ex\hbox{E}\kern-.125emX}}

\begin{document}

\title{Realizing Space-oriented Control in Smart Buildings via Word Embeddings
\thanks{}
}


\author{
    \IEEEauthorblockN{
        Hangli Ge\IEEEauthorrefmark{1}, Hiroaki Mori\IEEEauthorrefmark{2}, Yasuhira Chiba\IEEEauthorrefmark{2}, Noboru Koshizuka\IEEEauthorrefmark{1},\\
    }
    \IEEEauthorblockA{
       \IEEEauthorrefmark{1}\textit{Interfaculty Initiative in Information Studies}\\
        \IEEEauthorrefmark{2}Graduate School of Interdisciplinary Information Studies,\\ The University of Tokyo, Tokyo, Japan\\
        Email: \{hangli.ge, hiroaki.mori, yasuhira.chiba,noboru\}@koshizuka-lab.org
    }
}

\maketitle

\begin{abstract}
This paper presents a novel framework for implementing space-oriented control systems in smart buildings. In contrast to conventional device-oriented approaches, which often suffer from issues related to development efficiency and portability, our framework adopts a space-oriented paradigm that leverages natural language processing and word embedding techniques. The proposed framework features a chat-based graphical user interface (GUI) that converts natural language inputs into actionable OpenAI API calls, thereby enabling intuitive space level (e.g., room) control within smart environments. To support efficient embedding-based search and metadata retrieval, the framework integrates a vector database powered by Elasticsearch. This ensures the accurate identification and invocation of appropriate smart building APIs. A prototype implementation has been tested in a smart building environment at the University of Tokyo, demonstrating the feasibility of the approach.
\end{abstract}



\section{Introduction}


Smart buildings, which aim to improve indoor comfort, convenience, energy efficiency, and security, are also key components in realizing smart cities that leverage ICT technologies to deliver better services to people. To date, various applications utilizing advanced IoT technologies and AI have been researched and developed to support smart building development, leading to the increased utilization of building-related data. Additionally, spatial information within buildings is being actively digitized and used through technologies such as digital twins, enabling automated control inside smart buildings. At the same time, the siloed development of smart building applications has become increasingly prevalent. This has raised concerns about the burden on developers, highlighting the need for improved application portability through the use of building operating systems (building OS) and ontologies.

Accordingly, the development of smart building applications must consider \textbf{two key aspects: maintaining usability from the perspective of building users, and ensuring portability from the perspective of developers}. In this paper, \textbf{we propose a novel control framework designed to be both user-friendly and developer-friendly}. This framework integrates natural language input into API calls using embedding representations. We constructed a database of smart building APIs, implemented a query system based on embedding similarity calculations, and developed a chatbot-style user interface built upon this foundation. Finally, we demonstrate the broad generalizability and high portability of both the developed application and the underlying theoretical framework.

\begin{figure}[t]
\centering
\includegraphics[height=2.5in, width=1\linewidth]{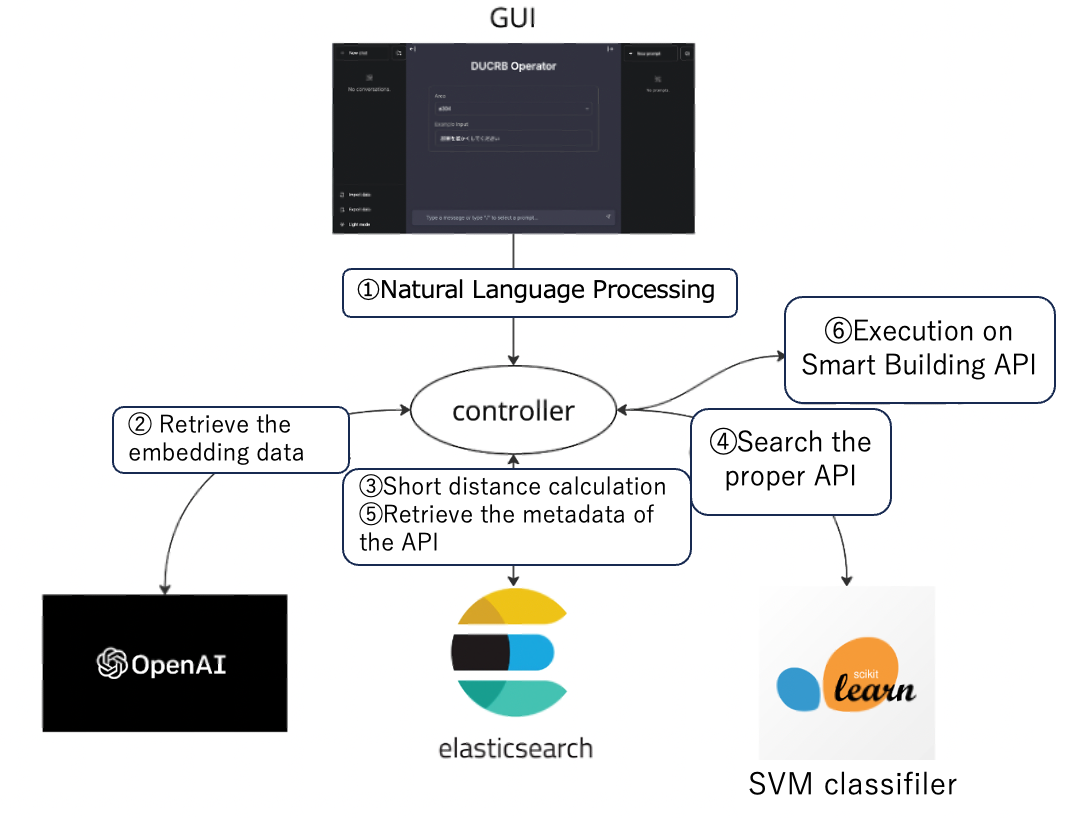}
\caption{Application Control Flow.}
\label{fig:flow}
\end{figure}

\begin{figure*}[t]
\centering
\includegraphics[height=2in, width=0.95\linewidth]{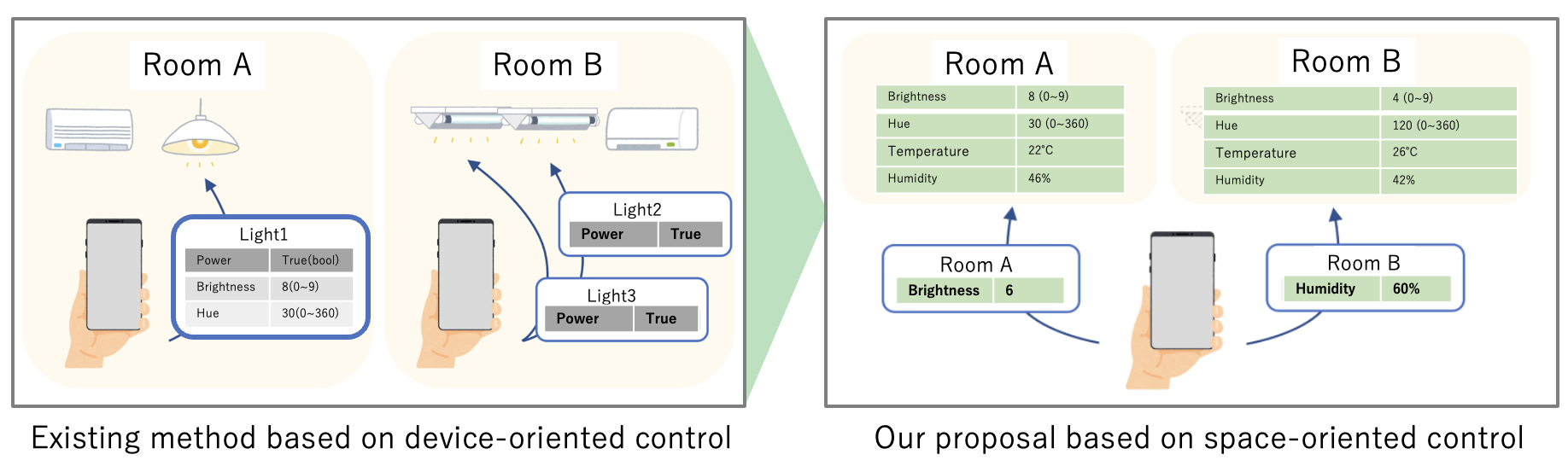}
\caption{Existing models VS our proposal}
\label{fig:over}
\end{figure*}
\section{Related Work}
In recent years, the growing momentum of AI research has extended into the smart building domain, resulting in the emergence of numerous AI-driven applications. Xia et al. employed reinforcement learning based on multitask learning for indoor temperature regulation, successfully reducing overall energy consumption \cite{xia2023reca}.
An et al. constructed a predictive model utilizing Gaussian processes, demonstrating that comparable energy-saving outcomes could be achieved with only seven days of training, in contrast to conventional approaches \cite{an2023clue}. Furthermore, several ontologies have been developed to represent the spatial configuration of buildings. For example, the SDB project developed an RDF-based metadata schema known as Brick \cite{balaji2016brick}, which demonstrates comprehensive metadata representation. Other ontologies such as BOT (Building Topology Ontology) \cite{rasmussen2020bot,ge2021applying} are also widely referenced in the architecture and building systems domains. In recent years, the integration of large language models (LLMs) into smart building systems has also gained traction. Li et al. developed an automated framework for constructing digital twins of smart buildings using LLMs \cite{li2023chattwin}. Similarly, a federated learning–based anomaly detection system tailored to the smart building context has been proposed \cite{sater2021federated}.

For the specific applications, Li et al. stated that a user-centric smart building design improves user satisfaction and well-being, emphasizing the need for a system that enables interaction between smart buildings and humans \cite{li2020review}. Jang et al. suggested the use of augmented reality by implementing holograms within smart buildings, implying potential applications in healthcare \cite{jang2018holosensor}.
Wang et al. enabled interaction with robots in virtual reality through facial expressions and blinking, making it possible to control IoT devices \cite{wang2019human}. Thus, various interaction modalities between humans and smart buildings are being explored, including visual interfaces, voice-based systems, and emotion recognition technologies.

Despite these advancements, several limitations have been identified. A critical concern is the lack of model portability. AI models trained for specific buildings, such as for energy optimization, often rely on unique metadata representations. Consequently, when such models are transferred to other buildings, additional preprocessing and adaptation steps are typically required, hindering practical deployment. Furthermore, the development of AI-based smart building applications necessitates the incorporation of building-specific physical and spatial information. However, the absence of a standardized schema for representing building information poses a significant challenge. Hence, the effective utilization of AIs in this context still requires sophisticated fine-tuning strategies.

In summary, while AI presents considerable potential for advancing smart building technologies, its practical implementation will remain limited in the absence of standardized data models capable of supporting interoperability and scalability across diverse building environments.

\section{Problem and Proposal}
\subsection{Problem}
As illustrated in Figure~\ref{fig:over}, existing method based on device-oriented controls, issue commands to individual devices. 
It refers to managing each individual device (e.g., thermostat, light switch, HVAC unit) in isolation, with control logic that is focused on the specific functionality of that device. It further causes the several key problems as follows:

\begin{enumerate}
    \item Lack of global optimization: Since each device acts independently, it’s difficult to coordinate actions across the entire building. For example, the air conditioning might run even when windows are open or occupancy is low in a room, leading to energy inefficiency. Smart buildings often require holistic, system-level optimization, something device-level logic can't deliver.
    \item Scalability or portability issues: As more devices are added to the building, the complexity of managing each one individually increases significantly. This leads to maintenance overhead and configuration challenges.
    \item Limited interoperability: Different devices may use different protocols or control standards. Device-level logic may not account for cross-vendor communication, making it hard to integrate heterogeneous systems.
    \item Difficulty in adapting to dynamic conditions: Smart buildings operate in dynamic environments (changing occupancy, weather, user preferences). Static device logic often fails to adapt, whereas system-level control (e.g. AI-based orchestration) can be more responsive.
    \item No support for goal-oriented control: Users or building managers may want to achieve high-level goals like: “Minimize energy cost” or “Maximize occupant comfort”. Device-oriented control can’t interpret or balance trade-offs between these goals.
\end{enumerate}

\subsection{Proposal}
We propose a novel control framework designed for space-oriented control that is both user-friendly and developer-friendly. \textbf{Based on this approach, users can manipulate the parameters of a space (e.g., a room) without needing to understand the specific parameters, locations, or functions of individual devices.} By incorporating ontologies into smart building development, it becomes possible to represent spatial configurations using standardized data models, thereby enhancing portability and interoperability across different environments.

Our space-oriented APIs define commands by specifying target values for the desired state of a space. The use of these APIs significantly improves application portability. Furthermore, we developed a chat-based graphical user interface (GUI) that translates natural language input into API calls using word embedding representations. This enables intuitive and user-friendly spatial control within smart building systems.

The overall architecture of the chat-based GUI was illustrated in Figure~\ref{fig:flow}. The details of the application’s control flow is listed as bellow:
\begin{enumerate}
    \item First, a natural language sentence is received as input from the GUI.
    \item The word embedding representation of the natural language sentence is then computed and retrieved.
    \item Based on the word embedding data, the system searches for the most similar example from a set of pre-prepared embedded examples in vector database. If the similarity score of the closest example falls below a certain threshold, the input is judged to be unrelated to building or spatial control, and the process is terminated.
    \item If the result from step (3) indicates that the input is related to building or spatial control, a classifier is used to determine which building API should be executed. This classifier takes an word embedding data as input and returns the corresponding API identifier category.
    \item Using the API identifier determined by the classifier, the system retrieves the API metadata from the database.
    \item Finally, the system extracts the endpoint, method, and body from the API metadata and executes the building API.
\end{enumerate}

\subsection{The front-end}
\begin{figure}[h]
\centering
\includegraphics[height=2in, width=0.9\linewidth]{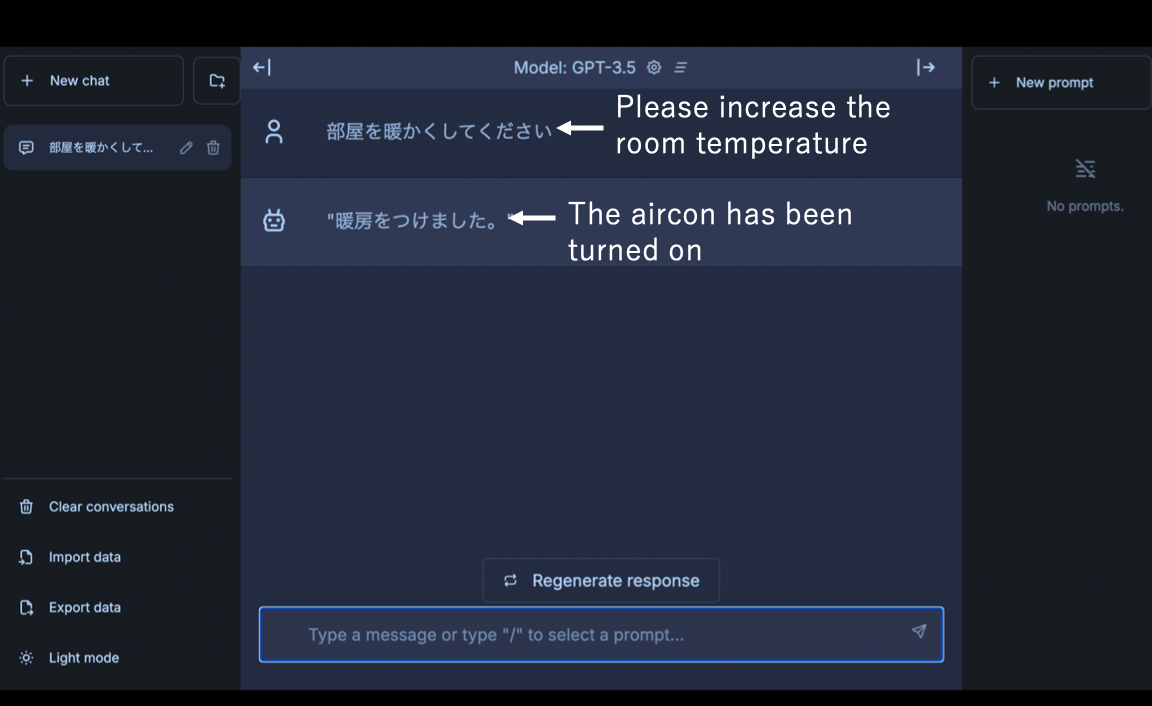}
\caption{Chat-based GUI}
\label{fig:aircon}
\end{figure}

The UI design was based on the open-source project chatbot-ui~\cite{chatbotui2024}, which is primarily implemented using TypeScript and Next.js. Moreover, TypeScript is supported by both the OpenAI Embeddings API client and the Elasticsearch client, as described later. Figure~\ref{fig:aircon} shows the chat-based UI. 

\subsection{The backend}
To compute embedding representations efficiently and cost-effectively, this study uses OpenAI’s publicly available API ~\cite{openai_embeddings2025}. Since it also supports TypeScript clients, the embedding API can be called directly from the controller. Among the two available third-generation embedding models by OpenAI, this application uses text-embedding-3-small, a lightweight and efficient model. The classifier for embedding representations was built using scikit-learn, a Python machine learning library~\cite{scikitlearn2025}. Python 3 was chosen as the backend language because it is also compatible with the Elasticsearch client, allowing both inference through the classifier and retrieval of API data from Elasticsearch indexes. The Flask web framework~\cite{flask2024} was used to develop a lightweight backend web server. The versions of the programming languages and tools used are shown in Table~\ref{tab:tools_versions}.
\begin{table}[h]
\centering
\begin{tabular}{|l|l|}
\hline
\textbf{Programming Language / Tool} & \textbf{Version} \\
\hline
Python3 & 3.10 \\
TypeScript & 4.9.5 \\
npm & 10.5.0 \\
Next.js & 13.2.4 \\
Elasticsearch client & 8.15.1 \\
scikit-learn & 1.5.2 \\
OpenAI client & 4.72.0 \\
Flask & 3.1.0 \\
\hline
\end{tabular}
\caption{Programming Languages and Tools with Versions}
\label{tab:tools_versions}
\end{table}


API metadata is represented in JSON format. A vector database was adopted due to the need to handle embedding data. These databases specialize in storing and processing vector embeddings and support high-speed vector search and data retrieval. Elasticsearch was chosen for its superior performance scores, multi-modal support, extensive documentation, and large user base. Even though API metadata (endpoint, method, body) are stored as plain strings rather than vectors, integrating them into Elasticsearch helps avoid the complexity and maintenance overhead of managing separate databases.

\begin{lstlisting}[
    caption=Mapping example embeddings in Elasticsearch,
    label=lst:mapping_example,
    basicstyle=\ttfamily\scriptsize, % ← 文字サイズ (\scriptsize や \tiny)
    aboveskip=4pt, belowskip=4pt    % ← 前後余白も縮めたい場合
]
{
  "mappings": {
    "properties": {
      "apiId": { "type": "text" },
      "order": { "type": "text" },
      "embedding": { "type": "dense_vector" }
    }
  }
}
\end{lstlisting}



\section{Experiment}

\begin{figure}
\centering
\includegraphics[height=2in, width=0.8\linewidth]{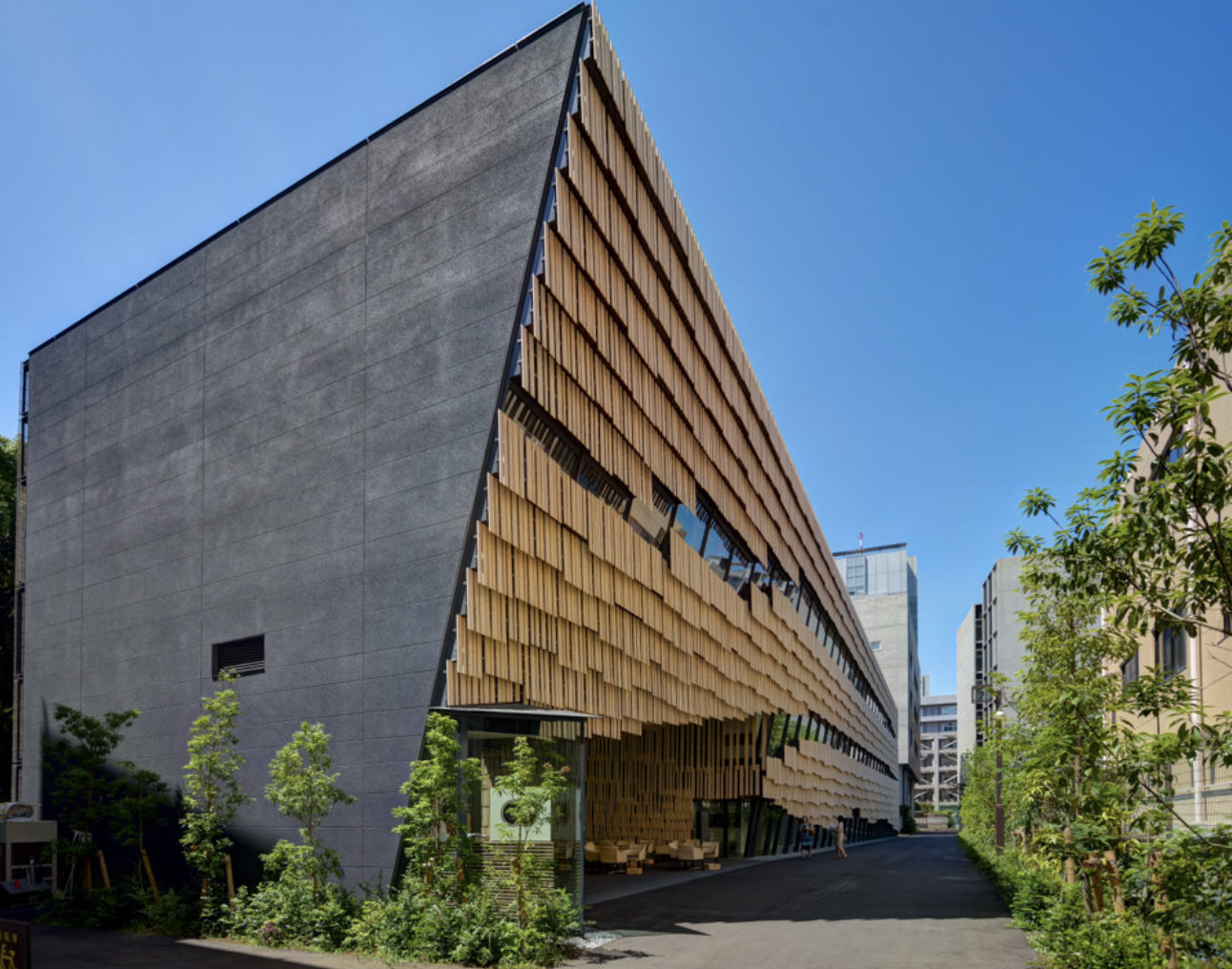}
\caption{Chat-based GUI}
\label{fig:build}
\end{figure}

A prototype was developed to operate in the student rooms and hall of the Daiwa Ubiquitous Computing Research Building (Figure~\ref{fig:build}) on the University of Tokyo’s Hongo Campus ~\cite{choi2020iot,taniguchi2022detecting, ge2022accurate, ge2024time}, which served as the target smart building.

Other types of functionality, such as transaction-based APIs can also be incorporated within the broader framework of smart building APIs. As listed in  ~\autoref{lst:trans}, consider a scenario where a user inputs a simple command such as "I'm leaving leaving the office". In this case, the system is expected to perform three actions via APIs: turning off the air conditioning in the user’s room, switching off the lights, and calling the elevator to the floor where the room is located.

\begin{lstlisting}[
    caption=Example of control transaction,
    label=lst:trans,
    basicstyle=\ttfamily\scriptsize, % ← 文字サイズ (\scriptsize や \tiny)
    aboveskip=4pt, belowskip=4pt    % ← 前後余白も縮めたい場合
]
[
  {
    "api_id": "leave_office",
    "transaction": [
      {
        "method": "PUT",
        "endpoint": "http://x.x.x.x/api/airconditioner",
        "body": "{\"ac_id\": \"A305\", \"on_off\": \"off\"}"
      },
      {
        "method": "PUT",
        "endpoint": "http://x.x.x.x/api/airconditioner",
        "body": "{\"light_id\": \"A305\", \"on_off\": \"off\"}"
      },
      {
        "method": "PUT",
        "endpoint": "http://x.x.x.x/api/elevator",
        "body": "{\"operation\": \"3fdown\"}"
      }
    ]
  }
]
\end{lstlisting}

\section{Conclusions}
In this study, we developed a novel space-oriented control framework that converts natural language input into API calls using word embedding representations, with the goal of enabling intuitive control for both end users and developers. In the current implementation, the database of building control APIs was constructed manually, and the prototype focuses on two rooms. For future work, integrating API metadata into building schemas based on ontologies is expected to be a promising direction for supporting control across an entire building. By extending such schemas to include entities that represent APIs, it would become possible to issue a single query that retrieves not only the available APIs but also the associated spaces and objects. This capability would significantly facilitate the implementation of the data model proposed in this study.

\bibliographystyle{ieeetr}
\bibliography{refs}

\end{document}